\documentclass[preprint,12pt]{elsarticle}

\usepackage{amssymb}

\usepackage{listings}
\usepackage{balance}
\usepackage{algorithm}
\usepackage[noend]{algpseudocode}
\algnewcommand\And{\textbf{ and }}
\algnewcommand\Not{\textbf{not }}
\usepackage{graphicx}
\usepackage[caption=false]{subfig}
\usepackage{color}
\usepackage{multirow}
\usepackage{textcomp}

\usepackage{pbox}
\usepackage{footnote}
\usepackage{url}

\usepackage{stackengine,graphicx,xcolor}
\usepackage{tikz}

\usepackage{booktabs} 
\newcommand{\ra}[1]{\renewcommand{\arraystretch}{#1}}

\definecolor{listinggray}{gray}{0.9}
\definecolor{lbcolor}{rgb}{0.9,0.9,0.9}

\journal{Parallel Computing}

\makeatletter
\def\ps@pprintTitle{%
 \let\@oddhead\@empty
 \let\@evenhead\@empty
 \def\@oddfoot{}%
 \let\@evenfoot\@oddfoot}
\makeatother

\begin{document}

\begin{frontmatter}



\title{DMR API: Improving the Cluster Productivity by Turning Applications into Malleable\tnoteref{lab1}}
\tnotetext[lab1]{This is an extended version of a previous work presented in the ICPP'17 workshop P2S2~\cite{Iserte2017}}
\tnotetext[]{DOI: 10.1016/j.parco.2018.07.006. \textcopyright 2018 Elsevier. This manuscript version is made available under the CC-BY-NC-ND 4.0 license http://creativecommons.org/licenses/by-nc-nd/4.0/}


\author[uji]{Sergio Iserte\corref{lab2}}
\ead{siserte@uji.es}
\author[uji]{Rafael Mayo}
\ead{mayo@uji.es}
\author[uji]{Enrique S. Quintana-Ort\'i}
\ead{quintana@uji.es}
\author[bsc]{Vicen\c{c} Beltran}
\ead{vbeltran@bsc.es}
\author[bsc]{Antonio J. Pe\~na}
\ead{antonio.pena@bsc.es}
\cortext[lab2]{Corresponding author}
\address[uji]{Universitat Jaume I (UJI), Castell\'o de la Plana, Spain}
\address[bsc]{Barcelona Supercomputing Center (BSC)}

\begin{abstract}
Adaptive workloads can change on--the--fly the configuration of their jobs, in terms of number of processes.
In order to carry out these job reconfigurations, we have designed a methodology which enables a job to communicate with the resource manager and, through the runtime, to change its number of MPI ranks.
The collaboration between both the workload manager---aware of the
queue of jobs and the resource allocation---and the parallel
runtime---able to transparently handle the processes and the program
data---is crucial for our throughput-aware malleability methodology.
Hence, when a job triggers a reconfiguration, the resource manager will check the cluster status and return an action: an expansion, if there are spare resources; a shrink, if queued jobs can be initiated; or none, if no change can improve the global productivity.
In this paper, we describe the internals of our framework and how it
is capable of reducing the global workload completion time along with providing a smarter usage of the underlying resources.
For this purpose, we present a thorough study of the adaptive workloads processing by showing the detailed behavior of our framework in representative experiments and the low overhead that our reconfiguration involves.
\end{abstract}


\end{frontmatter}


\section{Introduction}

In production HPC facilities, applications run on shared computers where hundreds or thousands of other applications are competing for the same resources. 
In this scenario, applications are submitted to the system with the shape of parallel jobs that conform the current workload of a system.
Adapting the workload to the infrastructure can render considerable improvements in resource utilization and global throughput.
A potential approach to obtain the desired adaptivity consists in applying \textit{dynamic job reconfiguration}, which devises resource usage to be potentially changed at execution time.

The adaption of the workload to the target infrastructure brings benefits to both system administrators and end users.
While administrators would like to see the throughput rate increased and a smarter resource utilization by the applications, 
end-users are the direct beneficiaries of the scale adaptivity, since
it removes strict resource requirements at submission time.
Although this may prevent the application from being executed in the
shortest time, users generally experience a faster completion time (waiting plus execution time).

In order to dynamically adapt a workload to the infrastructure, we need two main tools: (i) a resource manager system (RMS) capable of modifying the resources assigned to a job; and (ii) a parallel runtime to rescale an application.
In our solution, we have connected these components by developing a communication layer between the RMS and the runtime.

In this work we enhance the Slurm Workload Manager~\cite{Yoo2003} to achieve fair dynamic resource assignment while maximizing the cluster throughput.
We select Slurm\footnote{\url{http://slurm.schedmd.com}} because it is open-source, portable, and highly scalable.
Moreover, it is one of the most widely-adopted RMSs in the Top500 List\footnote{\url{http://www.top500.org}}.

To exploit a collection of distributed resources, the vast majority of the scientific applications that run on high performance clusters use the Message Passing Interface (MPI),
either directly or on top of programming models or libraries leveraging MPI underneath. 
\textit{Reconfiguration} is possible in MPI applications via the MPI spawning functionality.

The direct use of MPI to handle migrations, however, requires considerable effort from
skilled software developers in order to manage the whole set of data transfers among processes in different communicators.
For this purpose, we benefit from the recently-incorporated offload semantics of the OmpSs programming model~\cite{Sainz2015} to ease the malleability process and data redistribution.
In addition, we adapt the Nanos++ OmpSs runtime to interact with Slurm.
We improve the \texttt{Nanos++} runtime to reconfigure MPI jobs and establish direct communication
with the RMS. For that, applications will expose ``reconfiguring
points'' where, signaled by the RMS, the runtime will assist to resize the job on--the--fly.
We highlight that, although we benefit from the OmpSs infrastructure and semantics for job reconfiguration, 
our proposal may well be leveraged as a specific-purpose library, and applications using this solution are free to implement on-node parallelism using other programming models such as OpenMP.

%

In summary, the main contribution of this paper is a mechanism to accomplish MPI malleability, based on existing components (MPI, OmpSs, and Slurm) that enhances resource usage in order to produce higher global throughput in terms of executed jobs per unit of time. 
To that extent, we propose (1) an extension of the OmpSs offload mechanism to deal with dynamic reconfiguration; 
(2) a reconfiguration policy for the RMS to decide whether a job must be expanded or shrunk; 
and (3) a communication protocol for the runtime to interact with the
RMS, based on application-level Application Programming Interface (API) calls. 
Last, (4) we also provide an extensive evaluation of the framework that demonstrates the benefits of our workload-aware approach.

This article extends the previous work in~\cite{Iserte2017} by providing an extensive overhead study of the scheduling and resize times for reconfiguring jobs.
Furthermore, we have improved the workload execution analysis with a comparison job--to--job when these have been launched in the fixed and the flexible version. 
We focus our attention on how each job behaves when studying its individual time of waiting, executing and completing.

The rest of this paper is structured as follows:
Section~\ref{sec:related} discusses related work.
Section~\ref{sec:method} presents an overview of the proposed methodology.
Sections~\ref{sec:slurm} and~\ref{sec:nanos} present the extensions developed in
the Slurm RMS and the Nanos++ runtime in order to support our programming model proposal discussed in Section~\ref{sec:pm}.
Section~\ref{sec:results} evaluates and analyzes malleability in a production environment.
Finally, Section~\ref{sec:conclusions} outlines the conclusions and discusses future work.

\section{Related Work}
\label{sec:related}
In general, a job (application) may be classified in one of the following types: \textit{rigid}, \textit{moldable}, \textit{malleable} and \textit{evolving}~\cite{Feitelson1996}.
These classes depend on the number of concurrent processes during the execution of a job, so that we collapse them into two categories:
\begin{itemize}
 \item Fixed: The number of parallel processes remains constant during the execution (rigid and moldable applications).
 \item Flexible: The number of processes can be reconfigured on--the--fly, allowing distinct numbers of parallel processes in different parts of the execution (malleable and evolving applications) or job malleability.
 This action is known as \textit{dynamic reconfiguration}.
\end{itemize}


The first steps toward malleability targeted \textit{shared-memory} systems exploiting the flexibility of applications. 
In~\cite{Padhye1996} the authors leveraged moldability together with preemptive policies, such as equipartitioning and folding.
These policies can interrupt active jobs in order to redistribute processors among the pending jobs.

Checkpointing mechanisms have been used in the past to save the application state and resume its execution with a different number of processes, or simply to migrate the execution to other processes.
The work in~\cite{ElMaghraoui2009} explores how malleability can be
used in checkpoint\slash restart applications. 
There, a checkpoint--and--reconfigure mechanism is leveraged to restart applications with a different number of processes from data stored in checkpoint
files. Storing and loading checkpoint files, however, poses a nonnegligible overhead versus runtime data redistribution.


In~\cite{Lemarinier2016}, the authors aim at malleability using two different approaches:
in the first approach they use traditional checkpoint-restart mechanisms, leveraging the library Scalable Checkpoint\slash Restart for MPI (SCR)~\cite{scr}, to relaunch a job with a new number of processes after saving the state.
The second approach is based on the User Level Failure Migration (ULFM) MPI standard proposal for fault-tolerance~\cite{ulfm}. For this purpose, the authors cause abortions in the processes in order to use the shrink-recovery mechanism implemented in the library, and then, resume the execution in a new number of processes.

A resizing mechanism based on CHARM++ is presented in~\cite{Gupta}.
The authors of that work demonstrate the benefits of resizing a job in terms of both performance and throughput, but they do not address the data redistribution problem during the resizing.

The authors of~\cite{Ribeiro2013} rely on an MPI implementation called EasyGrid AMS in order to adjust automatically the size of a job.
Another similar approach is found in~\cite{Martin2015}, where a performance-aware framework based on the Flex-MPI library~\cite{Martin2013} is presented.
That work leverages job reconfiguration in order to expand\slash shrink a
job targeting execution performance.
For that purpose, the framework monitors execution, predicts future performance, and balances the load.

In the literature we can also find several works that combine malleability with resource management.
ReSHAPE~\cite{Sudarsan2007} integrates job reconfiguration techniques with job scheduling in a framework that also considers the current performance of the execution.
Complementary research using this framework analyzes its impact on individual performance and throughput in small workloads~\cite{Sudarsan2009a,Sudarsan2009}.
That solution, however, requires all applications in the cluster to be
specifically-developed to be flexible under the ReSHAPE framework.
In a more recent work, they present a more in-depth study discussing the ReSHAPE behavior with a workload of 120 jobs~\cite{Sudarsan2016}.

An additional important contribution is~\cite{Prabhakaran2015}, where a batch system with adaptive scheduling is presented.
The authors in this paper enable the communication between the RMS Torque/Maui and Charm++ as a parallel runtime.
Charm++ applications are presented as automatically malleable thanks to checkpointing. 

Compared with previous work, we present a holistic throughput-oriented reconfiguration
mechanism based on existing software components that is compatible with unmodified non-malleable applications. 
Furthermore, in contrast with previous studies, we
configure our workloads not only leveraging synthetic applications.
\section{Methodology Overview}
\label{sec:method}
Slurm exposes an API that may be used by external software agents.
We use this API from the Nanos++ OmpSs runtime in order to design the job resize mechanism.
Thus, Slurm's API allows us to resize a job following the next steps:
\begin{itemize}
\item{Job A has to be expanded}
\begin{enumerate}
\item Submit a new job B with a dependency on the initial job A.
Job B requests the number of nodes NB to be added to job A.
\item Update job B, setting its number of nodes to~0.
This produces a set of NB allocated nodes which are not attached to any job.
\item Cancel job B.
\item Update job A and set its number of nodes to NA+NB.
\end{enumerate}

\item{Job A has to be shrunk}
\begin{enumerate}
\item Update job A, setting the new number of nodes to the final size (NA is updated).
\end{enumerate}
\end{itemize}

After these steps, Slurm's \textit{environment variables} for job A are updated. 
These commands have no effect on the status of the running job, and the user remains responsible for any malleability process and data redistribution.

\begin{figure}
  \includegraphics[clip,width=\columnwidth]{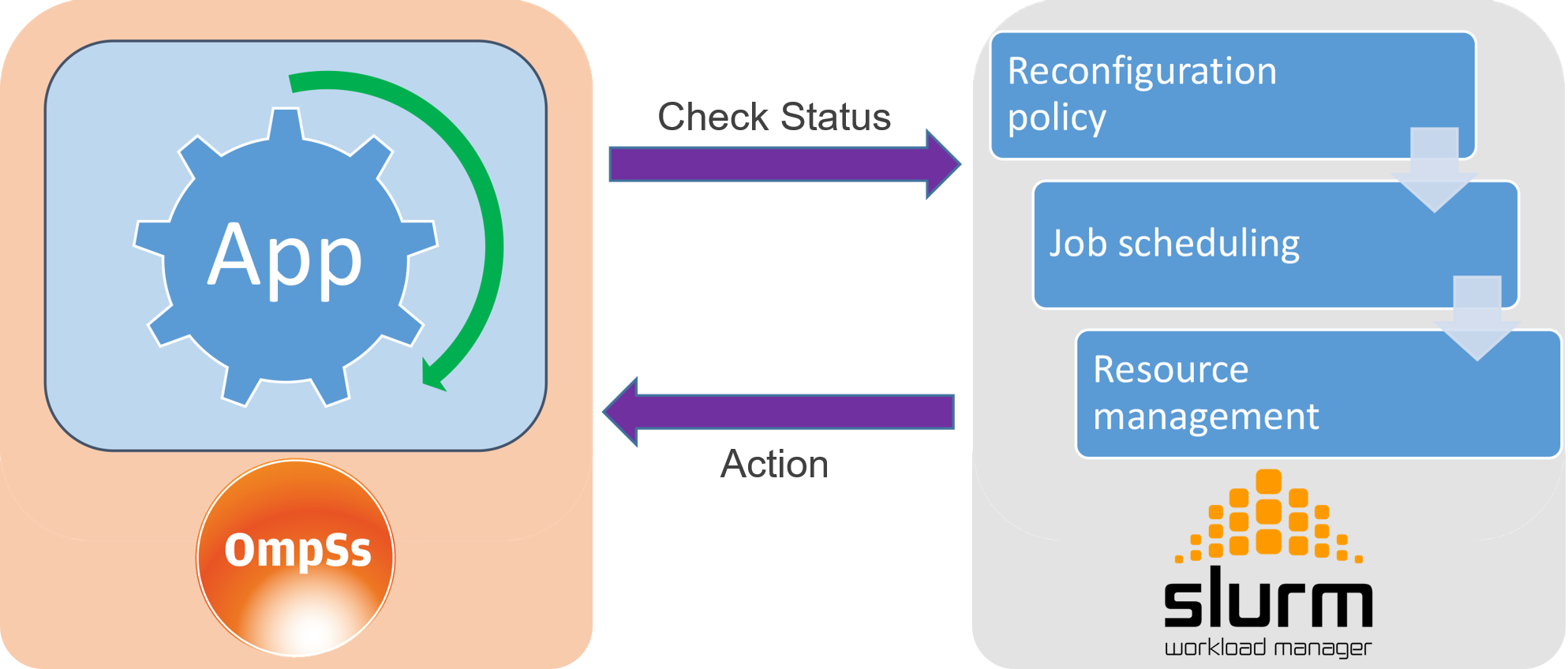}
\caption{Scheme of the interaction between the RMS and the runtime.}\label{fig:overview}
\end{figure}
The framework we leverage consists of two main components: the RMS
and the programming model runtime (see Figure~\ref{fig:overview}).
The RMS is aware of the resource utilization and the queue of pending jobs.
When an application is in execution, it periodically contacts the RMS, through the runtime, communicating its rescaling willingness (to expand or shrink the current number of allocated nodes).
The RMS inspects the global status of the system to decide whether to initiate any rescaling action, and communicates this decision to the runtime. 
If the framework determines that a rescale action is due, the RMS, the runtime, and the application will collaborate to continue the execution
of the application scaled to a different number of nodes (MPI processes).

\section{Slurm Reconfiguration Policy}
\label{sec:slurm}
We designed and developed a resource selection plug-in responsible for reconfiguration decisions.
This plug-in realizes a node selection policy featuring three modes that accommodate three degrees of scheduling freedom.

\subsection{Request an Action}
Applications are allowed to ``strongly suggest'' a specific action.
For instance, to expand the job, the user could set the ``minimum'' number of requested nodes 
to a value that is greater than the number of allocated nodes.
However, Slurm will ultimately be responsible for granting
the operation according to the overall system status.

\subsection{Preferred Number of Nodes}
One of the parameters that applications can convey to the RMS is their \emph{preferred} number of nodes to execute a specific computational stage.
If the desired size corresponds to the current size, the RMS will return ``no action''.
If a ``preference'' is requested and there is no outstanding job in the queue, the expansion can be granted up to a specified ``maximum''.
Otherwise, if the desired value is different from the current allocation, the RMS will try to expand or shrink the job to the preferred number of nodes.

\subsection{Wide Optimization}
\label{subsubsec:optimization}
The cases not covered by the preceding methods are handled as follows:

\begin{itemize}
\item
\emph{A job is expanded} if there are sufficient available resources to
fulfill the new requirement of nodes and either
(1) there is no job pending for execution in the queue, or
(2) no pending job can be executed due to insufficient available resources.
By expanding the job, we can expect it to finish its execution earlier
and release the associated resources.

\item
\emph{A job is shrunk}
if there is any queued job that could be executed by taking this action. More jobs in execution should increase the global throughput.
Moreover, if the job is going to be shrunk, the queued job that has triggered the shrinking event will be assigned the maximum priority in order to foster its execution.
\end{itemize}

\section{Framework Design}
\label{sec:nanos}
We implemented the necessary logic in Nanos++ to reconfigure jobs
in tight cooperation with the RMS. In this section we discuss the
extended API and the resizing mechanisms.

\subsection{The Dynamic Management of Resources (DMR) API}
\label{subsubsec:dmr}
We designed the DMR API with two main functions:
\texttt{dmr\_check\_status} and its asynchronous version
\texttt{dmr\_icheck\_status}. These routines instruct the runtime
(Nanos++) to communicate with the RMS (Slurm) in order
to determine the resizing action to perform: ``expand'',
``shrink'', or ``no action''. 
The asynchronous counterpart schedules the next action for the next execution step, at the same time that the current step is executed.
Hence, by skipping the action scheduling stage, the communication overhead in that step is avoided.

In case an action is to be performed, these functions spawn a new set of processes and return an opaque handler.
This API is exposed by the runtime and it is intended to be used by applications.
These functions present the following input arguments.
\begin{itemize}
 \item Minimum number of processes to be resized to.
 \item Maximum number of processes. This prevents the application from growing beyond its scalability capabilities.
 \item Resizing factor (e.g., a factor of 2 will expand/shrink the number of processes to a value multiple/divisor of 2).
 \item Preferred number of processes.
\end{itemize}
The output arguments return the new number of nodes and an opaque handler
to be used in subsequent operations.

An additional mechanism implemented to attain a fair balance between performance and throughput is the ``checking inhibitor''.
This introduces a timeout during which the calls to the DMR API are ignored.
This knob is mainly intended to be leveraged in
iterative applications with short iteration intervals.
The inhibition period can be tuned by means of an environment variable.

\subsection{Automatic Job Reconfiguration} 
\label{subsubsec:reconfiguration}
The runtime will perform the following actions in order to leverage the Slurm resizing mechanisms (see
Section~\ref{sec:slurm}) by means of its external API.

\subsubsection{Expand}
A new \textit{resizer} job (RJ) is first submitted requesting the difference between the
current and total amount of desired nodes. 
This enables the original nodes to be reused.
There is a dependency relation between the RJ and the original job (OJ).
In order to facilitate complying with the RMS decisions, RJ is set to the maximum priority, facilitating its execution.

The runtime waits until JR moves from the ``pending'' to the ``running'' status.
If the waiting time reaches a threshold, RJ is canceled and the action
is aborted.
This situation may occur if the RMS assigns the available resources to a different job during the scheduling action.
This is more likely to occur in the asynchronous mode because an action then can experience some delay during which the status of the queue may change.
Once OJ is reallocated, the updated list of nodes is gathered and used in a call to \texttt{MPI\_Comm\_spawn} in order to create a new set of processes.

\subsubsection{Shrink}
The shrinking mechanism is slightly more complex than its expansion
counterpart because Slurm will have to kill all processes executing in the released nodes.
To prevent premature process termination, we need a synchronized workflow to guide the job shrinking.
Hence, the RMS sets a \textit{management node} in charge of receiving an acknowledgment from all other processes.
These ACKs will signal that they finished their tasks and the node is ready to be released.


After a scheduling is complete, the DMR call returns the expand--shrink action to be performed and the resulting number of nodes.
The application is responsible for performing the appropriate actions
using our proposed programming model as described in Section~\ref{sec:pm}.

\section{Programming Model}
\label{sec:pm}




In this section we review our programming model approach to address
dynamic reconfiguration coordinated by the RMS. The programmability of
our solution benefits from relying on the OmpSs offload semantics versus directly using MPI. 

\subsection{Benefits of the OmpSs Offload Semantics}
\label{subsec:benefits}

To showcase the benefits of the OmpSs offload semantics, we review the
specific simple case of migration. 
This analysis allows us to focus on the fundamental differences between programming models because it does not involve data redistribution among a different number of nodes (which is of similar complexity in both models). 

\paragraph{MPI Migration}
Listing~1 contains an excerpt of pseudo-code directly using MPI calls. 
In this case, we assume some mechanism is available to determine the new node list in line 17.

\begin{lstlisting}[float,caption=Pseudo-code of job reconfiguration using bare MPI., label=code:mpi, captionpos=b]
void main(int argc, char **argv) {
  ...
  int t = 0;
  MPI_Comm_get_parent(&parentComm);
  if (parentComm == MPI_COMM_NULL) {
    init(data);
  } else {
    MPI_Recv(parentComm, data, myRank);
    MPI_Recv(parentComm, &t, myRank);
  }
  compute(data, t);
  ...
}

void compute(data, t0) {
  for (t=t0; t<timesteps; t++) {
    nodeList = get_new_nodelist_somehow();
    if (nodelist != NULL) {
      MPI_Comm_spawn(myapp.bin, nodeList, &newComm);
      MPI_Send(newComm, data, myRank);
      MPI_Send(newComm, t, myRank);
      exit(0);
    }
    compute_iter(data, t);
  }
}
\end{lstlisting}





\paragraph{OmpSs-based Migration}
The same functionality is attained in Listing~\ref{code:ompss} by leveraging our proposal on top of the OmpSs offload semantics. 
This includes a call to our extended API in line 11.
At a glance, our proposal exposes higher-level semantics, increasing code expressiveness and programming productivity.
In addition, communication with the RMS is implicitly established in the call to the runtime in line 11, which pursues an increase in overall system resource utilization.
Data transfers are managed by the runtime with the directive in line 13.
Moreover, at this point, the initial processes terminates, letting the
execution of ``compute'' in line 14 continue in the processes of the
new communicator identified by ``handler''.

\begin{lstlisting}[float,caption=Pseudo-code of job reconfiguration using OmpSs., label=code:ompss, captionpos=b]
void main(void) {
  ...
  int t = 0;
  init(data);
  compute(data, t);
  ...
}

void compute(data, t0) {
  for (t=t0; t<timesteps; t++) {
    action = dmr_check_status(..., &newNnodes, &handler);
    if (action) {
      #pragma omp task inout(data) onto(handler, myRank)
      compute(data, t)
    } else
      compute_iter(data);
  }
}
\end{lstlisting}

\subsection{A Complete Example}
\label{subsec:ompss}

\begin{figure}
\subfloat[Expand.]{%
  \includegraphics[clip,width=0.5\columnwidth,trim={0cm 8.2cm 0cm 0cm}]{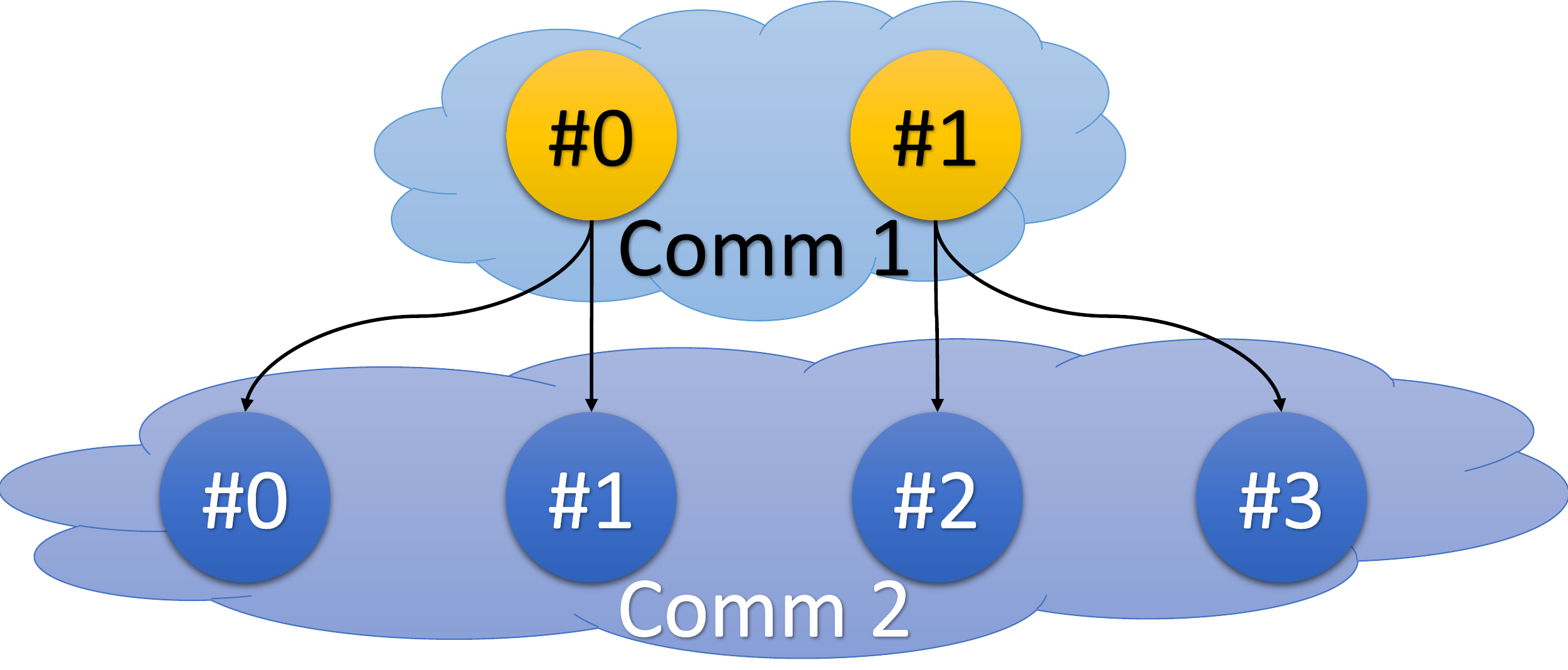}\label{subfig:expand}
}
\subfloat[Shrink.]{%
  \includegraphics[clip,width=0.5\columnwidth,trim={0cm 8.2cm 0cm 0cm}]{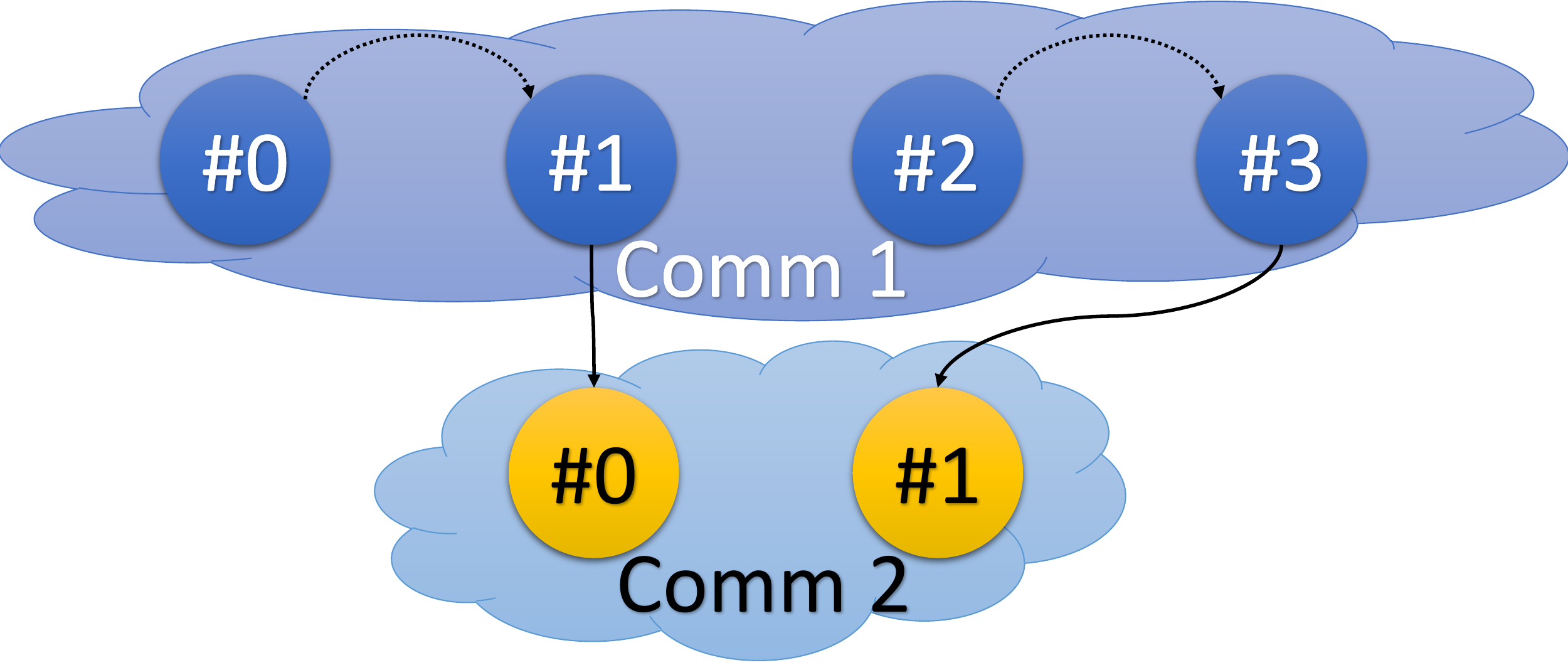}\label{subfig:shrink}
}
\caption{Data transfers.}\label{fig:transfers}
\end{figure}

The excerpt in Listing~\ref{code:malleable} is derived from that
showcased in Section~\ref{subsec:benefits} to discuss malleability.
In this case the application must drive the task redistribution according to
the resizing action.
The mapping \texttt{factor} indicates the number of processes in the
current set that are mapped to the processes in the new configuration (see Figure~\ref{fig:transfers}).
This example implements homogeneous distributions, where we
always resize to a multiple or a divisor of the current number of processes.
Our model, however, supports arbitrary distributions.

\begin{lstlisting}[float,caption=Pseudo-code of a malleable application., label=code:malleable, captionpos=b]
void compute(data, t0) {
  for (t=t0; t<timesteps; t++) {
    action = dmr_check_status(..., &newNnodes, &handler);
    if (!action)
      compute_iter(data);
    else {  
      if (action == "expand") {
        factor = newNnodes / worldRanks;
        for (i=0; i<factor; i++) {
          dest = myRank * factor + i;
          subdata = part_data(factor, data);
          #pragma omp task inout(subdata) onto(handler,dest)
          compute(subdata, t);
        }  // End for
      } else if (action == "shrink") {
        factor = worldRanks / newNnodes;
        sender = (myRank % factor) < (factor - 1);
        if (sender) {
          dst = factor * (myRank / factor + 1) - 1;
          MPI_Isend(comm, data, dst);
        } else {  // Receiver
          for (i=1; i<=factor; i++) {
            src = myRank - factor + i;	
            MPI_Irecv(comm, &alldata, src);
          }  // End for
        }  // End if (sender)
        MPI_Waitall();
        if (!sender) {
          dest = myRank / factor;
          #pragma omp task inout(alldata) onto(handler,dest)
          compute(alldata, t);
        }  // End if (!sender)
      } else error();  // End if (action == ...)
    }  // End if (action)
  }  // End for
}  // End compute()
\end{lstlisting}



For the ``expand'' action (line 8),
the original processes must partition the dataset.
For instance, in Figure~\ref{subfig:expand}, the processes split the dataset into two subsets, mapping each half to a process in the new configuration.
The data transfers are performed by the runtime according to the information
included in the task offloading directive (line 12).

The ``shrink'' action, on the other hand, involves preliminary explicit data movement.
The processes in the original set are grouped into ``senders'' and
``receivers''.
This initial data movement is illustrated in the example in Figure~\ref{subfig:shrink}.

\section{Experimental Results} \label{sec:results}
In this section we evaluate the our implementation of the proposed
framework based on the DMR API via a set of workloads, fixed and flexible, composed of three applications: Conjugate Gradient (CG), Jacobi and N-body.
These applications were turned into malleable in order to study malleability in a production environment.
Furthermore, we developed a synthetic application in order to perform
an isolated overhead assessment.

The description of the applications and their behavior, as well as a thorough preliminary study of all the features implemented in the DMR API can be found in~\cite{Iserte2017}.
Table~\ref{tab:apps-conf} summarizes the reconfiguration parameters for each application.

\begin{table}
\caption{Configuration parameters for the applications}
\label{tab:apps-conf}
\centering
\ra{1} 
\begin{tabular}{rccccc} 
\toprule
& & \multicolumn{3}{c}{Number of Processes} & \\ \cline{3-5}
Application & Iterations &  Minimum & Maximum & Preferred & Scheduling period\\ \midrule
FS & 25 & 1 & 20 & - & -\\
CG & 10000 & 2 & 32 & 8 & 15 seconds\\
Jacobi & 10000 & 2  & 32 & 8 & 15 seconds\\
N-body & 25 & 1 & 16 & 1 &  -\\
\bottomrule
\end{tabular}
\end{table}

\subsection{Workload Configuration}
\label{subsec:workload}
The workloads were generated using the statistical model proposed by Feitelson~\cite{Feitelson1996}, which characterizes rigid jobs based on observations from logs of actual cluster workloads. 
For our purpose, we leverage the model customizing the following 2 parameters:
\begin{itemize}
	\item \texttt{Jobs}: Number of jobs to be launched.
	\item \texttt{Arrival}: Inter-arrival times of jobs modeled using a Poisson distribution of factor 10, which will prevent from receiving bursts of jobs while preserving a realistic job arrival pattern.	
\end{itemize}

\subsection{Platform}
Our evaluation was performed on the Marenostrum Supercomputer at {\em Barcelona Supercomputing Center}.
Each compute node in this facility is equipped with two 8-core Intel Xeon E5-2670 processors running at 2.6~GHz with 128~GB of RAM.
The nodes are connected via an InfiniBand Mellanox FDR10 network.
For the software stack we used MPICH 3.2, OmpSs 15.06, and Slurm 15.08. 

Slurm was configured with the \textit{backfill} job scheduling policy.
Furthermore, we also enabled job priorities with the policy \textit{multifactor}.
Both were configured with default values.

\subsection{Reconfiguration Scheduling Performance Evaluation}
We next analyze the overhead of using our framework to enable malleability.
For this purpose, we used the synthetic application Flexible Sleep (FS), configured to perform 2 steps and to transfer 1~GB of data during the reconfiguration.
The idea is that each job executes an iteration, then it contacts with the RMS, and resumes the execution in the second step with the new configuration of processes.

\begin{figure}
\subfloat[Scheduling  time.]{%
  \includegraphics[clip,width=0.5\columnwidth,trim={4.5cm 2.5cm 6.5cm 2.5cm}]{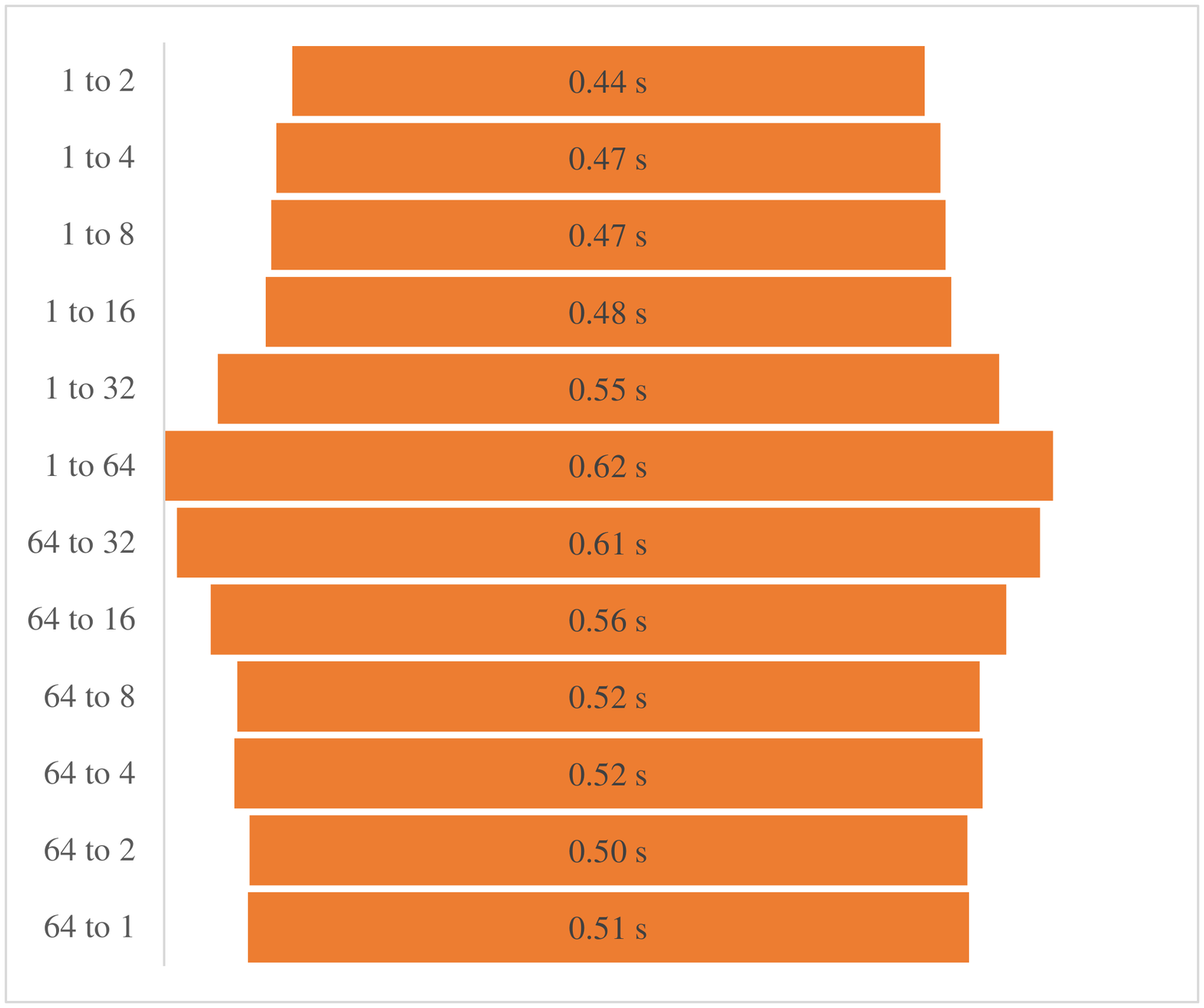}\label{subfig:schedTimes}
}
\subfloat[Resize time.]{%
  \includegraphics[clip,width=0.5\columnwidth,trim={4.5cm 2.5cm 6.5cm 2.5cm}]{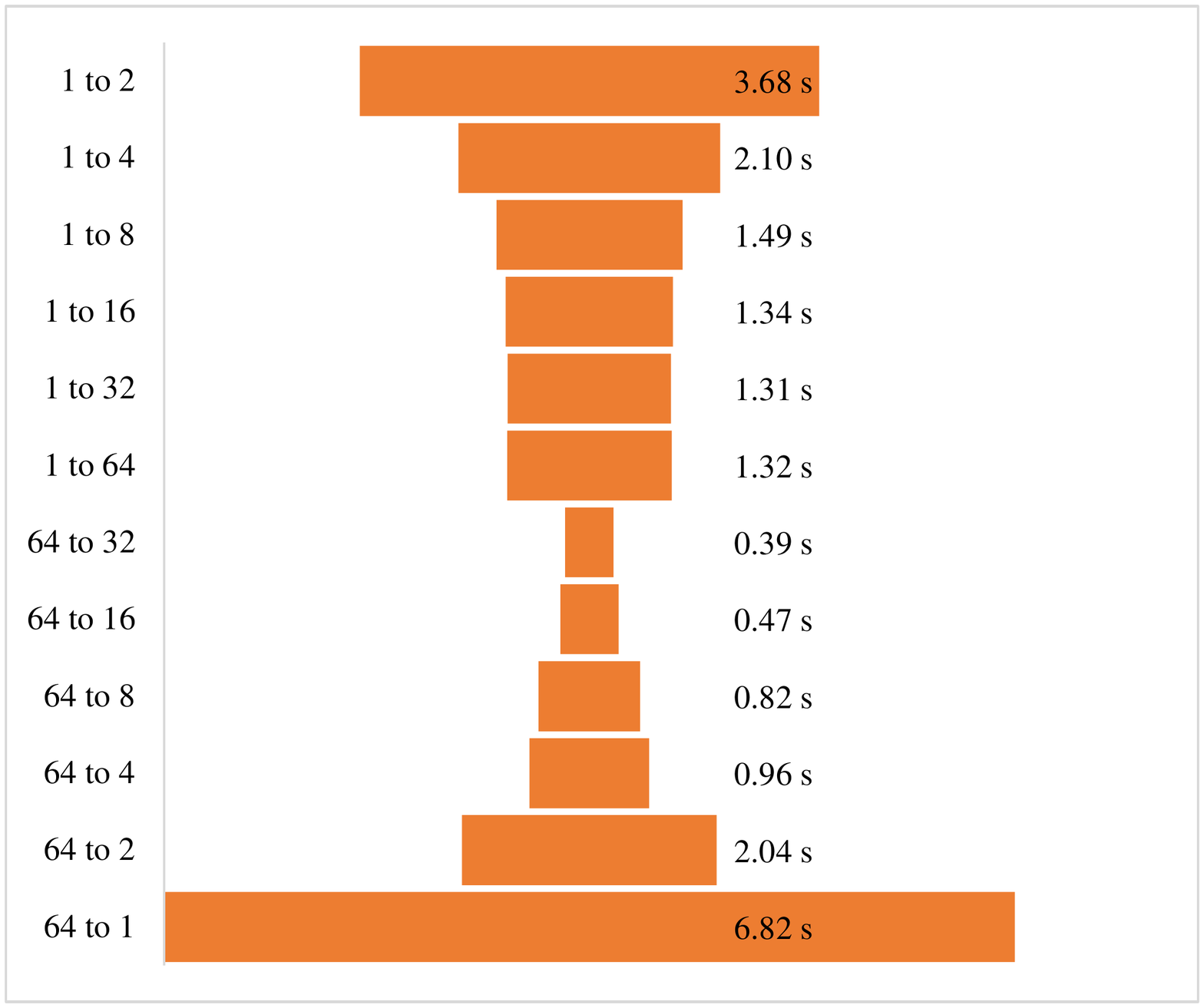}\label{subfig:resizTimes}
}
\caption{Time needed to reconfigure from/to processes.}\label{fig:reconfTimes}
\end{figure}

Figure~\ref{fig:reconfTimes} shows the average time of 10 executions for each reconfiguration.
On the left (a) we can see the times taken by the RMS to determine an action (scheduling time).
From top to bottom, the first half of the chart depicts the expansions, while the second half, the shrinks.
The chart reveals a slight increment in the scheduling time when more nodes are involved in the process.

The chart in Figure~\ref{fig:reconfTimes}(b) shows the time needed to
perform the necessary transfers among processes.
Two interesting behaviors are appreciated in this figure:
\begin{itemize}
 \item The more processes involved in the reconfiguration, the shorter resize time.
This is because the chunks of data are smaller and the time needed to transfer them concurrently is lower (compare the time between 1 to 2 and 64 to 32 processes).
 \item Shrinks involve much more synchronization among processes and the greater the difference in the number of processes is, the more time is needed to synchronize all of them.
\end{itemize}

 We have also studied the overhead in a real workload execution.  Table~\ref{tab:actions} reports the statistics collected for a 400-job workload.
 The table is divided in three parts and shows the actions taken and their run time during the workload execution in both synchronous and asynchronous scheduling.
 
 When no action is performed, the time to decide is negligible (average time and standard deviation of ``no action'').
 The time increases when the RMS performs an action because of the
 scheduling process itself and the reconfiguration operations performed by the runtime. 
 
 The first two rows of the second and the third part provide
 information about the number of reconfigurations scheduled per workload and per job.
 We can see that the synchronous version schedules fewer
 reconfigurations and not all of these jobs are expected to be resized.
 Moreover, since we are processing workloads with many queued jobs, running jobs are likely to be shrunk in favor of the pending. 
 
 The table also demonstrates the negative effect of a timeout during an expansion.
 This effect is shown in the asynchronous scheduling column with the ``maximum'', ``average'' and ``standard deviation'' values.
 In addition to the maximum time taken by the runtime to assert the expanding operation, these timeouts reveal a non-negligible dispersion in the duration values for the ``expand'' action.
 In fact, having such a high standard deviation turns the average time little representative.

  \begin{table}
 \caption{Analysis of the actions performed by the framework in a 400-job workload.}
 \label{tab:actions}
 \centering
 \begin{tabular}{cccc} 
 & & Synchronous & Asynchronous\\ \toprule
 \multicolumn{1}{ c }{\multirow{4}{*}{\parbox{1.5cm}{No\\ Action}}} &
 \multicolumn{1}{ c }{Minimum Time (s)} & 0.0010 & 0.0003\\
 \multicolumn{1}{ c  }{}                        &
 \multicolumn{1}{ c }{Maximum Time (s)} & 0.2078 & 0.1140\\
 \multicolumn{1}{ c  }{}                        &
 \multicolumn{1}{ c }{Average Time (s)} & 0.0094 & 0.0137\\
 \multicolumn{1}{ c  }{}                        &
 \multicolumn{1}{ c }{Standard Deviation (s)} & 0.0102 & 0.0112\\ \cmidrule{2-4}

 \multicolumn{1}{c}{\multirow{6}{*}{\parbox{1.5cm}{Action Expand}}} &
 \multicolumn{1}{c}{Quantity} & 50 & 107\\
 \multicolumn{1}{c}{}                        &
 \multicolumn{1}{c}{Actions/Job} & 0.125 & 0.267\\
 \multicolumn{1}{c}{}                        &
 \multicolumn{1}{c}{Minimum Time (s)} & 0.367 & 0.366\\
 \multicolumn{1}{c}{}                        &
 \multicolumn{1}{c}{Maximum Time (s)} & 0.530 & 40.418\\
 \multicolumn{1}{c }{}                        &
 \multicolumn{1}{c}{Average Time (s)} & 0.423 & 8.820\\
 \multicolumn{1}{c }{}                        &
 \multicolumn{1}{c}{Standard Deviation (s)} & 0.146 & 12.688\\ \cmidrule{2-4}
 
 \multicolumn{1}{c}{\multirow{6}{*}{\parbox{1.5cm}{Action Shrink}}} &
 \multicolumn{1}{c}{Quantity} & 194 & 303\\
 \multicolumn{1}{c}{}                        &
 \multicolumn{1}{c}{Actions/Job} & 0.485 & 0.757\\
 \multicolumn{1}{c}{}                        &
 \multicolumn{1}{c}{Minimum Time (s)} & 0.233 & 0.334\\
 \multicolumn{1}{c}{}                        &
 \multicolumn{1}{c}{Maximum Time (s)} & 0.541 & 0.555\\
 \multicolumn{1}{c}{}                        &
 \multicolumn{1}{c}{Average Time (s)} & 0.425 & 0.422\\
 \multicolumn{1}{c}{}                        &
 \multicolumn{1}{c}{Standard Deviation (s)} & 0.498 & 0.049\\ \bottomrule
 \end{tabular}
 \end{table}
 
\subsection{Dismissing the Asynchronous Scheduling}
In a previous work the asynchronous mode was tested presenting worse results than the synchronous counterpart~\cite{Iserte2017}.
This section thoroughly evaluates and compares both methods showing the inappropriateness of the asynchronous scheduling for processing adaptive workloads.
Table~\ref{tab:aggregated} compares both modes, synchronous and asynchronous, in more detail analyzing their performance at cluster level and at job level.
The most remarkable aspect here is that the synchronous scheduling occupies almost all the resources during the complete executions (the low standard deviation reveals that the mean value is barely unchanged for all the sizes).
Moreover, the asynchronous mode still presents a higher utilization rate than the configuration without flexible jobs.
However, the high standard deviation means that the utilization is not as regular as in the synchronous case.
In fact, this result hides a low average utilization for small workloads (as we already reported in this subsection, the small workloads performed worst) compared with a high  average for large workloads, similar to the synchronous scenario.

\begin{table}
\caption{Cluster and job measures of the 400-job workloads.}
\label{tab:aggregated}
\centering
\begin{tabular}{lrccc}
\toprule
\multicolumn{2}{c}{Cluster Measures} & Fixed & Synchronous & Asynchronous \\ \midrule
\multirow{2}{*} {Resources utilization} & Avg. (\%) & 83.607& 93.909& 86.687 \\
& Std. (\%) & 5.353&1.012 &  8.735 \\ \midrule
\multicolumn{2}{c}{Per Job Measures} & Fixed & Synchronous & Asynchronous \\ \midrule
\multirow{2}{*} {Waiting time gain} & Avg. (\%) & - &  27.980& 30.575\\
& Std. (\%) & -& 12.124 & 17.282\\
\multirow{2}{*} {Execution time gain} & Avg. (\%) & - &-58.482 & -97.294 \\
& Std. (\%) &- & 26.731 & 34.378\\
\multirow{2}{*} {Completion time gain} & Avg. (\%) & - & 12.786&  7.799\\
& Std. (\%) & -& 4.083& 5.548\\
\bottomrule
\end{tabular}
\end{table}
 
The three last columns offer information about timing measures: the wait-time of a job before entering execution, the execution time of the job, and the difference of time from the job submission to its finalization (completion).
Malleability provides an important reduction of the wait-time in both modes for all the sizes.
This is because the resource manager can shrink a job in execution in favor of a queued one.

With respect to the execution time, we experience a high degradation in the performance of each individual job.
For the synchronous scheduling, the negative gain of around a 50\% is closely related to the fact that the application scales linearly.
Thus, halving the resources produces a proportional reduction in performance.
In the asynchronous scenario, the degradation is even more pronounced.
The high standard deviation means that the performance of all the jobs
is not equally impacted; in fact, the jobs in the small workloads are the most affected instances (as shown in~\cite{Iserte2017}).

Finally, the global job time (completion time) is what places
malleability as an interesting feature, especially the synchronous
scheduling that completes the jobs, on average, 12\% earlier than the traditional scenario.

Since this test reveals no benefit from using asynchronous scheduling,
the rest of the experiments will exclusively use the synchronous mode.

\subsection{Throughput Evaluation}
For our throughput evaluation we generated workloads of 50, 100, 200 and 400 jobs for both versions, fixed and flexible.
Each workload is composed of a set of randomly-sorted jobs (with a fixed seed) which instantiate one of the three non-synthetic applications: Conjugate Gradient (CG), Jacobi and N-body.
For every malleable job the \texttt{shrink-expand factor} was set to 2.
The job submission of each application is launched with its ``maximum'' value, reflecting the user-preferred scenario of a fast execution. 

\begin{figure}
\centering
  \includegraphics[clip,width=0.9\columnwidth, trim={2.1cm 10cm 1.9cm 11.35cm}]{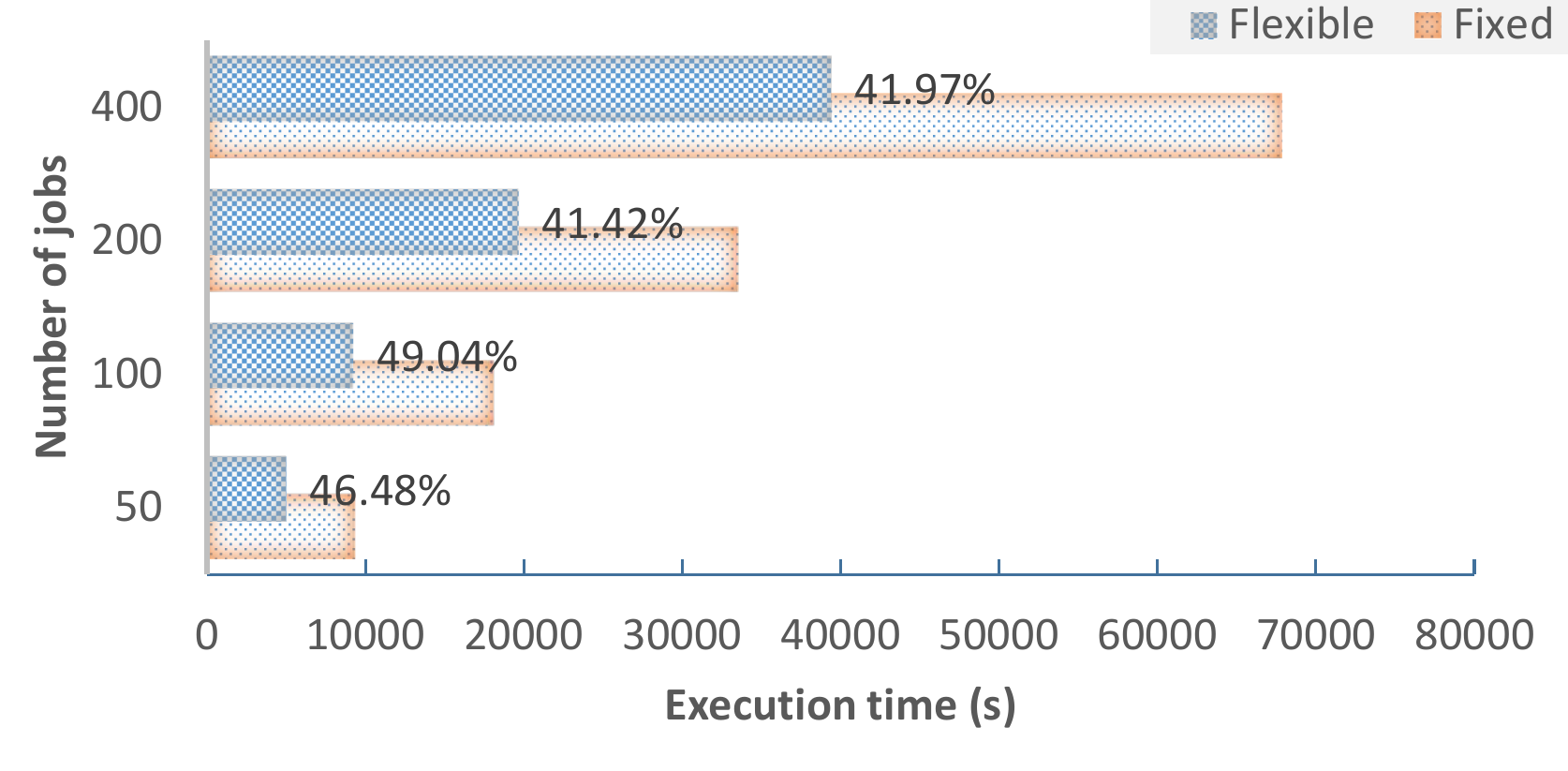}
\caption{Workloads execution times (bars) and the gain of flexible workloads (bar labels).}\label{fig:exps}
\end{figure}

Figure~\ref{fig:exps} depicts the execution time of each workload size comparing both configuration options: fixed and flexible.
The labels at the end of the ``flexible'' bars report the gain compared with the fixed version.
Table~\ref{tab:summary} details the measures extracted from the executions.
In the first column, we compare the average resource utilization for fixed and flexible workloads.
This rate corresponds to the average time when a node has been allocated by a job compared to the workload completion time.
These results indicate that the flexible workloads reduce the allocation of nodes around 30\%, offering more possibilities for queued jobs. 

\begin{table}
\caption{Summary of the averaged measures from all the workloads.}
\label{tab:summary}
\centering 
\begin{tabular}{ccllll}
\toprule
\#Jobs & Version & \pbox{2cm}{Resource Utilization Rate} & \pbox{2cm}{Job\\Waiting Time} & \pbox{2cm}{Job\\Execution Time} & \pbox{2cm}{Job\\Completion Time}\\ \midrule
\multirow{2}{*}{50} & Fixed & 98.71 \% & 4115.02 s. &  620.26 s. & 4735.28 s.\\
 & Flexible & 68.67 \% & 1359.92 s. & 900.3 s. & 2260.22 s.\\
\multirow{2}{*}{100} & Fixed & 97.39 \%&9750.34 s.& 586.64 s.&10336.98 s.\\
 & Flexible & 71.91 \%&2990.6 s.&858.16 s.&3848.76 s.\\
 \multirow{2}{*}{200} & Fixed & 98.38 \%&17466.2 s.&520.58 s.&17986.78 s.\\
 & Flexible & 73.54 \%&6856.8 s.&825.88 s.&7676.67 s.\\
 \multirow{2}{*}{400} & Fixed & 98.38 \%& 31788.39 s.&532.14 s.&32320.53 s.\\
 & Flexible & 73.54 \%&13861.03 s.& 843.19 s..& 14704.22 s.\\
\bottomrule
\end{tabular}
\end{table}

\begin{figure}
\centering
  \includegraphics[clip,width=0.9\columnwidth, trim={2.1cm 10cm 1.9cm 11.35cm}]{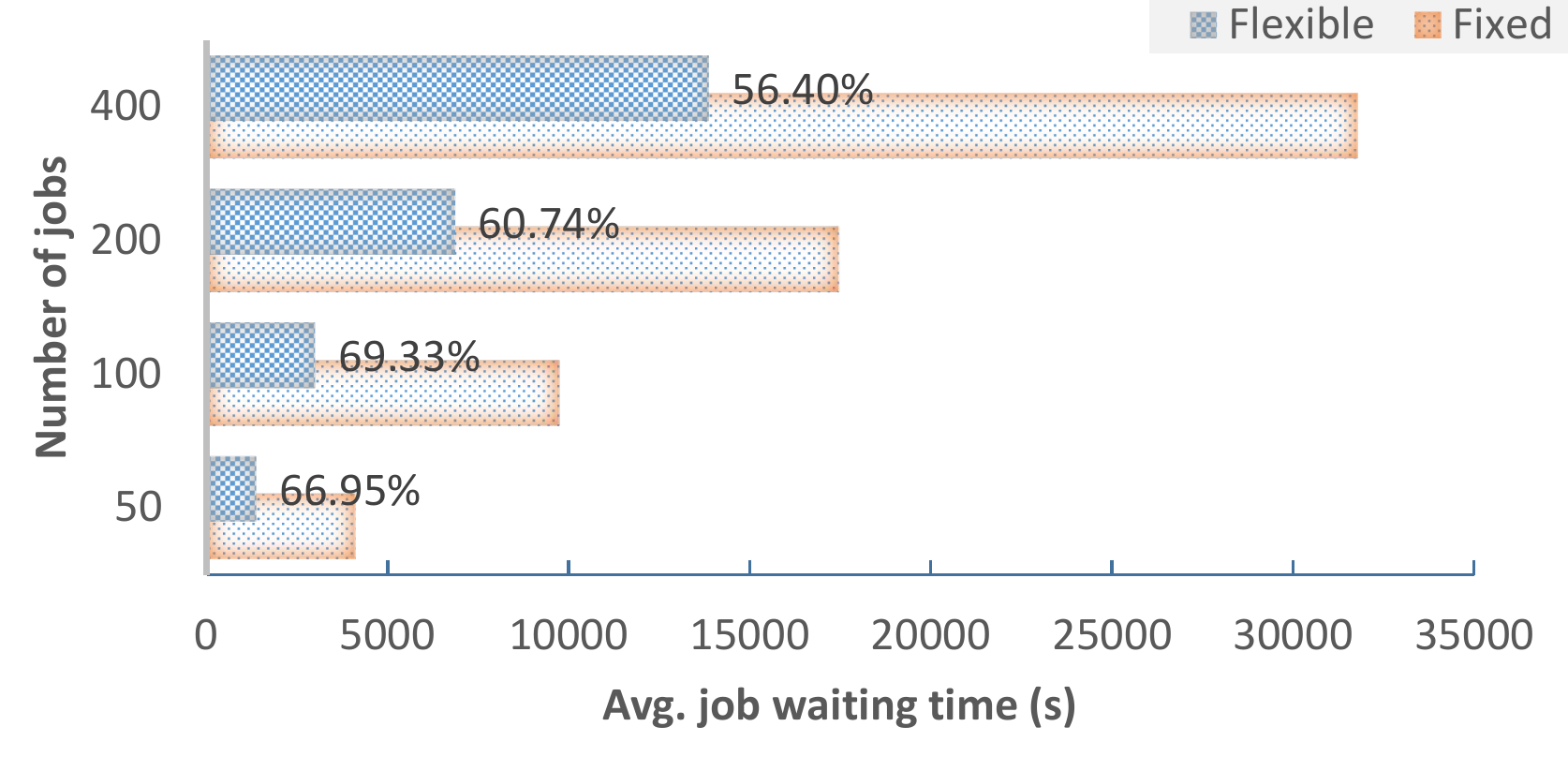}
\caption{Average waiting time for all the jobs of each workload (bars) and the gain of flexible workloads (bar labels).}\label{fig:waits}
\end{figure}

The second column of Table~\ref{tab:summary} shows the average waiting time of the jobs for each workload.
These times are illustrated in Figure~\ref{fig:waits}, together with the gain rate for flexible workloads.
The around 60\% reduction makes the job waiting time a crucial measure
to consider from the throughput perspective.
In fact, this time is the responsible for the reduction in the workload execution time.

The last two columns of Table~\ref{tab:summary} present two more aggregated measures of all the jobs in the workload:
The first one is the average execution time;
the second is this execution time plus the waiting time of the job, referred to as completion time.
The experiments show that jobs in the flexible workload are affected by the scale-down in their number of processes.
Nevertheless, this is compensated by the waiting time which benefits the completion time.

In order to understand the events during a workload execution, we have chosen the smallest workload to generate detailed charts and offer an in-depth analysis.

\begin{figure}
\centering
\begin{tikzpicture}
    \node[anchor=south west,inner sep=0] (image) at (0,0) {\includegraphics[clip,width=0.75\columnwidth]{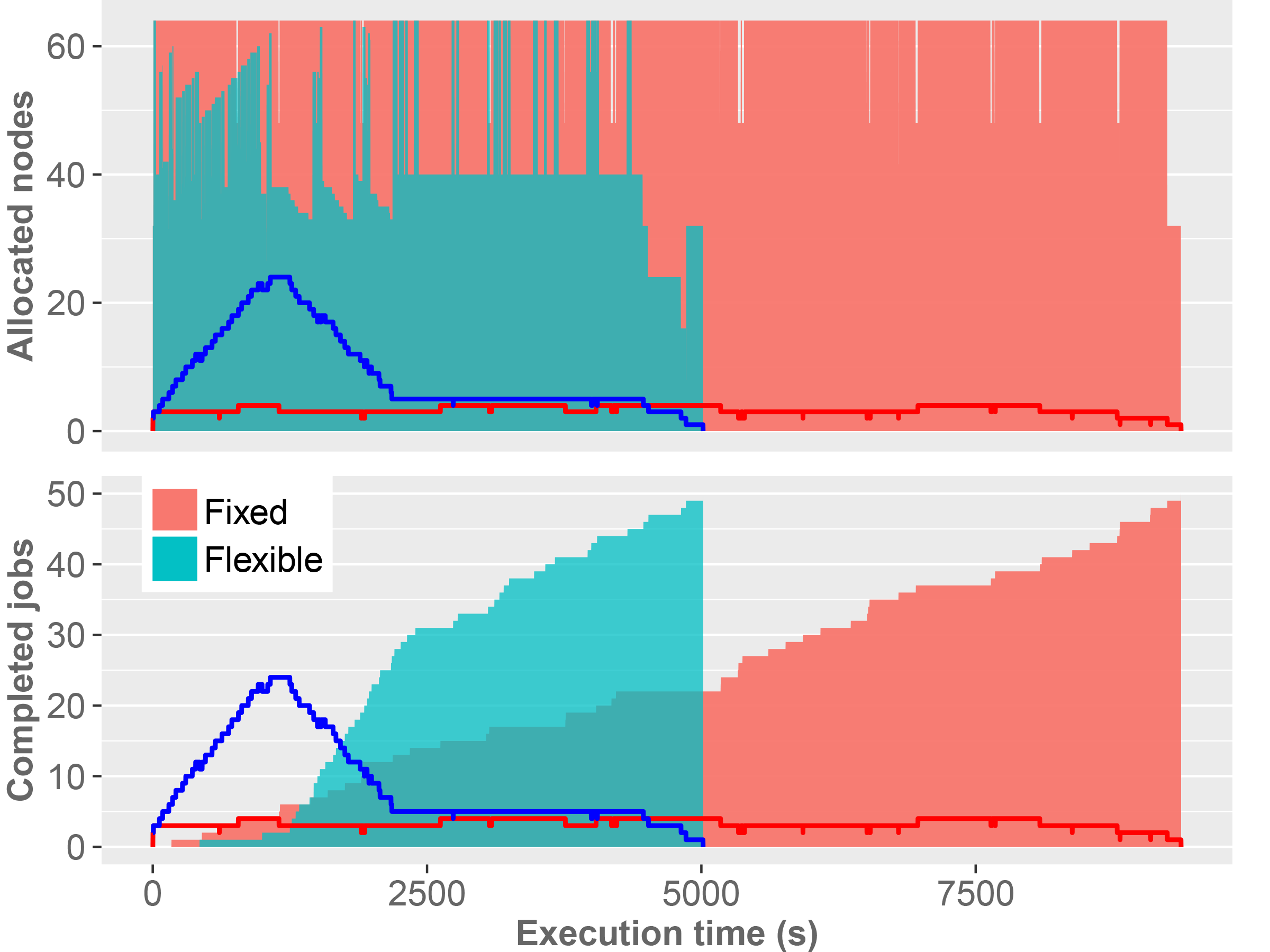}};
    \begin{scope}[x={(image.south east)},y={(image.north west)}]
        \draw[black,thick,rounded corners] (0.315,0.97) rectangle (0.5,0.555);
    \end{scope}   
\end{tikzpicture}
  \caption{Evolution in time for the 50-job workload. Blue and Red lines represent the running jobs for fixed and flexible policies.}\label{fig:alloc}
\end{figure}

The top and bottom plots in Figure~\ref{fig:alloc} represent the
evolution in time of the allocated resources and the number of
completed jobs, respectively.
The figure also shows the number of running jobs for fixed and flexible workloads (blue and red lines respectively).
These demonstrate that the flexible workload utilizes fewer resources; furthermore, there are more jobs running concurrently (top chart).
For both configurations, jobs are launched with the ``sweet spot''
number of processes (in terms of parallel efficiency); the fixed jobs obviously do not vary the amount of assigned resources, while in the flexible configuration, they are scaled-down as soon as possible.
This explains the reduction on the  utilization of resources.
For instance, in the second half of the flexible shape in Figure~\ref{fig:alloc} (marked area), we find a repetitive pattern in which there are 5 jobs in execution which allocate 40 nodes.
The next eligible job pending in the queue needs 32 nodes to start.
Therefore, unless one of the running jobs finishes, the pending job will not start and the allocation rate will not be higher.
When a job eventually finishes and releases 8 nodes, the scheduler initiates the job requesting 32 nodes.
Now, there are 64 allocated nodes (the green peaks in the chart); however, as the job prefers 8 processes, it will be scaled-down.
 
At the beginning of the trace in the bottom of Figure~\ref{fig:alloc},
the throughput of the fixed workload is higher than that of its counterpart; this occurs because the first jobs are completed earlier (they have been launched with the best-performance number of processes).
Meanwhile, in the flexible workload, many jobs are initiated (blue
line) and, as soon as they start to finish, the overall throughput experiences
a notable improvement.

\begin{figure}
\centering
  \includegraphics[clip,width=0.9\columnwidth]{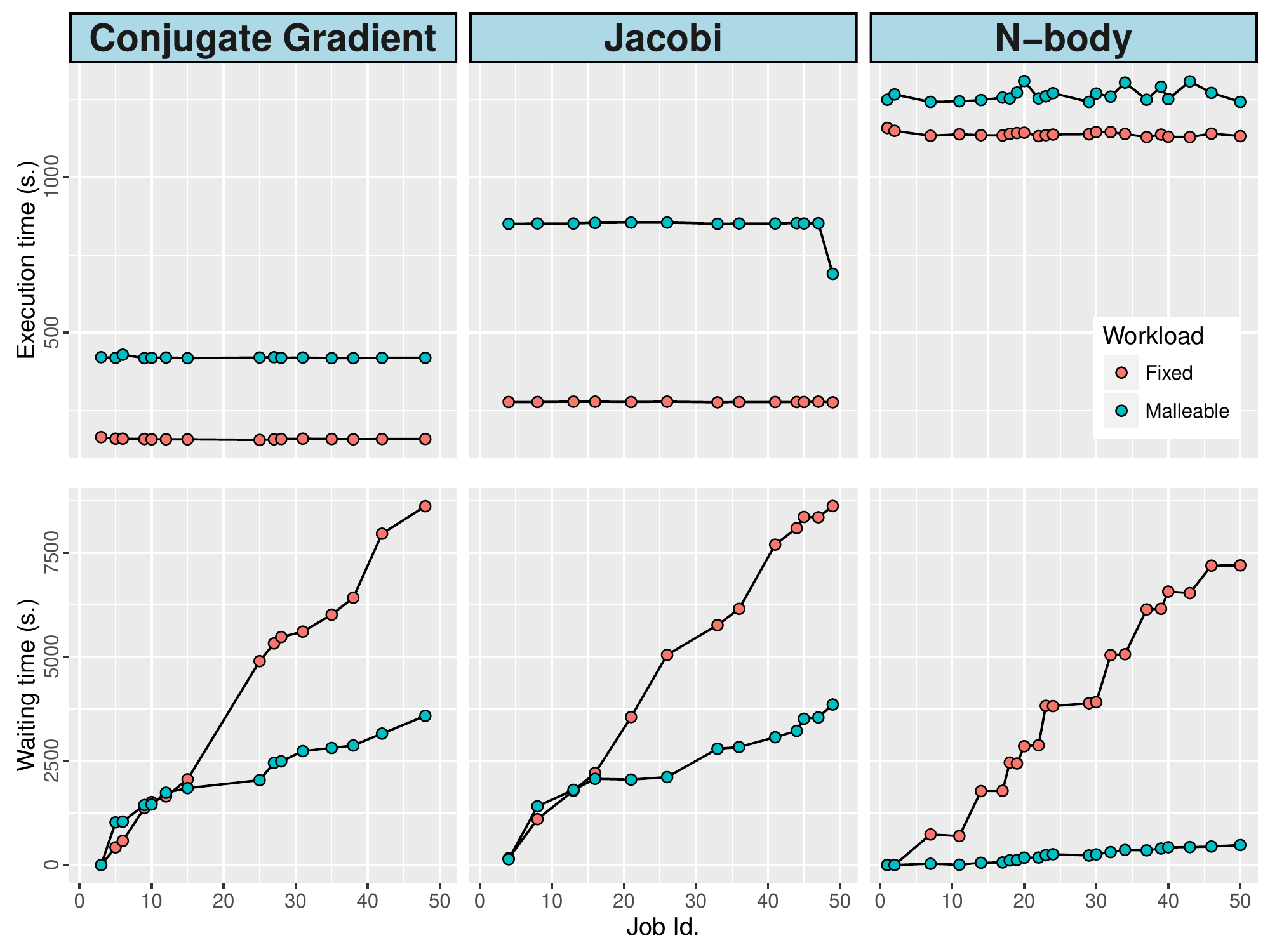}
  \caption{Execution (top) and waiting (bottom) times of each job grouped by applications (columns).}\label{fig:timesApps}
\end{figure}

Figure~\ref{fig:timesApps} depicts the execution and waiting time of each job grouped by application.
The execution time (top row of charts) increases in the flexible workload for all the cases.
As mentioned earlier, we are shrinking jobs to their preferred value as soon as these are initiated.
This implies a performance decrease because of a lack of resources.
However, there is a job that leverages the benefits of an expansion.
The last Jacobi job experiences a drop on its execution time, since it
has been expanded thanks to completed jobs that released their resources.

The row of charts at the bottom of the figure compares the waiting time of all the jobs for their fixed and flexible versions.
At the beginning there is no remarkable difference in the waiting time of both versions; however,
the emergence of new jobs continues and resources remain allocated for the running jobs in the fixed workload. The RMS cannot provide the means for draining faster the queue.
For this reason, queued jobs in the fixed workload, experience a considerable delay in their initiation.

\begin{figure}
\centering
  \includegraphics[clip,width=0.9\columnwidth]{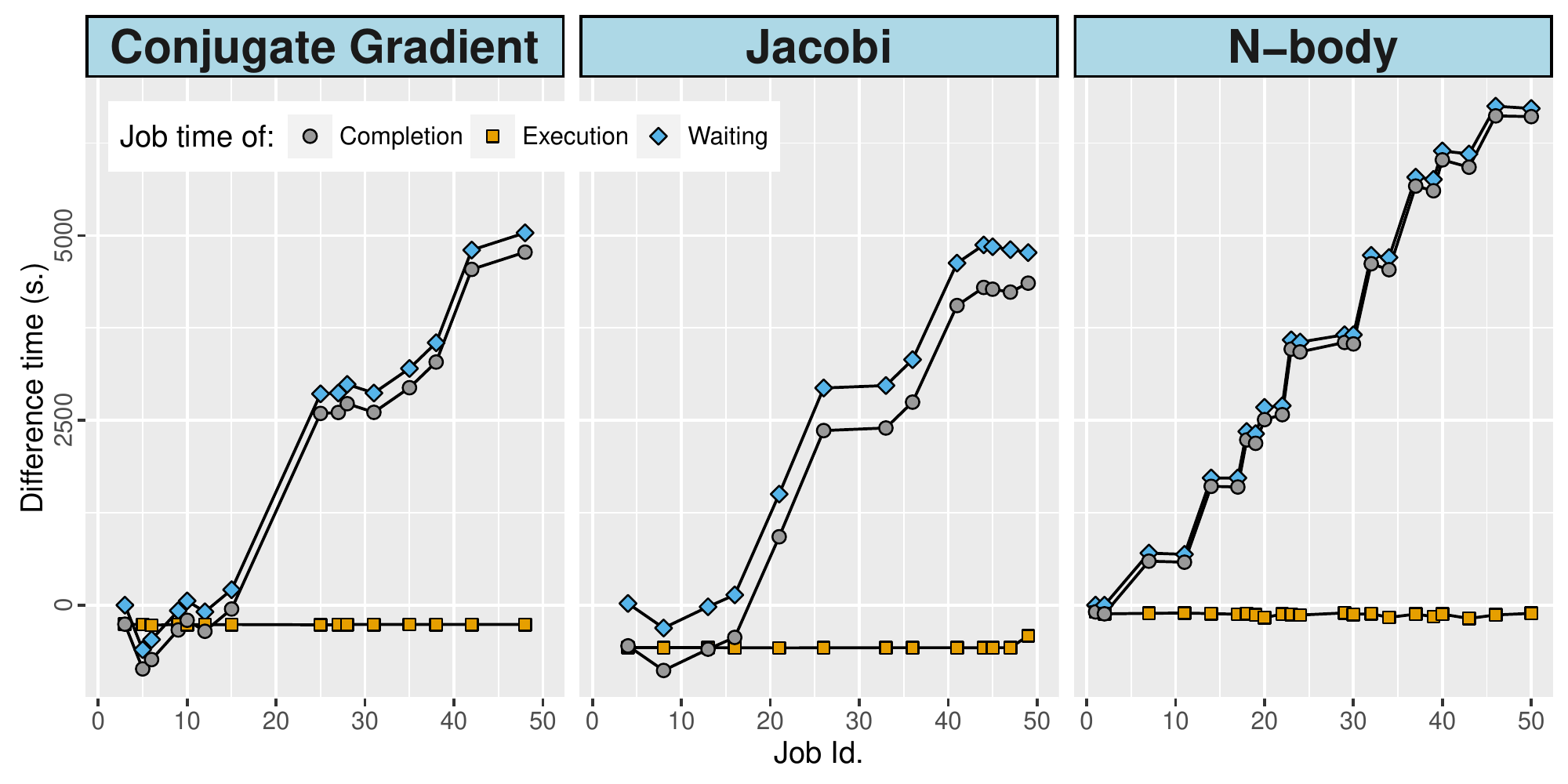}
  \caption{Time difference of the fixed vs.\ flexible executions for each job of their completion, execution and waiting time.}\label{fig:diffTimesApps}
\end{figure}

That difference of the fixed vs.\ flexible executions in the initiation is crucial for the completion time of the job, as shown in Figure~\ref{fig:diffTimesApps}.
This figure represents the difference in execution, waiting and completion time for each job grouped by application.
Again, the execution time remains below zero, what means that the difference is negative and the flexible workload performs slower.
Nevertheless, this small drawback is highly compensated by the waiting time.
As can be seen, completion difference time shows a heavy dependency on the waiting time,
making it the main responsible for reducing the individual completion time, and in turn, the high throughput obtained in the experiments.

\section{Conclusions and Future Work}
\label{sec:conclusions}
This paper improves the state--of--the--art in \textit{dynamic job reconfiguration} 
by targeting the global throughput of a high performance facility.
We benefit from already-existing first-class
software components to design our novel approach that introduces a dynamic reconfiguration mechanism for malleable jobs, composed of two modules: 
the runtime and the resource manager.
Those two elements collaborate in order to resize jobs on--the--fly to favor the global throughput of the system. 

As we prove in this paper, our approach can significantly improve resource utilization while, at the same time, reducing the wait-time for enqueued jobs, and decreasing the total execution time of workloads. 
Although this is achieved at the expense of a certain increase in the job execution time, we have reported that, depending on the scalability of the application, this drawback can be negligible.






\section*{Acknowledgments}
The authors would like to thank the anonymous reviewers for their
valuable and insightful comments that improved the quality of this
paper. This work is supported by the projects TIN2014-53495-R and
TIN2015-65316-P from MINECO and FEDER. This project has received
funding from the European Union's Horizon 2020 research and innovation
programme under the Marie Sklodowska Curie grant agreement No. 749516.
\balance

\bibliographystyle{elsarticle-num}
\bibliography{bib/icpp17}

\end{document}